\documentclass[twocolumn,showpacs,preprintnumbers,amsmath,amssymb]{revtex4-1}
\usepackage{graphicx}
\usepackage{textcomp}

\begin{document}
\title{Field-induced barrier transparency of Bloch waves in tight-binding lattices}
  \normalsize
\author{Stefano Longhi 
}
\address{Dipartimento di Fisica, Politecnico di Milano, Piazza L. da Vinci
32, I-20133 Milano, Italy}

%
\bigskip
\begin{abstract}
\noindent A rectangular potential barrier for a Bloch particle in a
tight-binding lattice is shown to become fully transparent by the
application of a strong ac field with appropriate amplitude and
frequency. Such a curious phenomenon bears some connection with the
 field-induced barrier transparency effect known for freely-moving
 particles scattered by an ac-driven rectangular barrier; however,
for a Bloch particle transparency is not related to a resonant
tunnneling process across the cycle-averaged oscillating potential
barrier, as for the freely-moving quantum particle. The phenomenon
of field-induced transparency is specifically discussed here for
photonic transport in waveguide arrays and demonstrated by full
numerical simulations of the paraxial (Schr\"{o}dinger) wave
equation beyond the tight-binding approximation.
\end{abstract}



\maketitle

\section{Introduction}
Coherent control of electronic, photonic or matter wave transport in
driven semiconductor superlattices, quantum dots, waveguide arrays
and optical lattices has received a great and continuous interest
over the past two decades \cite{tras1,tras2,tras3,tras4},
stimulating a wide number of experimental and theoretical
investigations in different physical systems
\cite{DL,CDT,vari,DellaValle,Longhi06,Arimondo}. Examples of
coherent control by the application of ac fields include, among
others, the coherent destruction of tunneling between two wells in a
bistable potential \cite{CDT} and the suppression of quantum
diffusion and dynamic localization in tight-binding lattices
\cite{DL}. Such phenomena have been demonstrated in a series of
recent experiments \cite{DellaValle,Longhi06,Arimondo}, mainly based
on transport of light waves or cold atoms in waveguide arrays or
optical lattices, where dephasing and many-body effects can be
neglected. Other interesting phenomena occur in the tunneling and
scattering processes of ac-driven free particles across potential
wells or barriers \cite{f1,f0,chaos1,chaos2}. In particular, a
freely-moving quantum particle can be resonantly transmitted across
a periodically-driven rectangular potential barrier, in spite the
probability of tunneling through the static potential barrier is
almost zero \cite{f1,f0}. This phenomenon, originally predicted for
quantum tunneling and referred to as field-induced barrier
transparency \cite{f1,f0}, is rather generic and can be similarly
observed for optical tunneling at modulated dielectric interfaces
\cite{f2}. In the accelerated (Kramers-Henneberger) reference frame,
the ac-driven barrier behaves like a periodically-shaken potential
barrier \cite{f1,chaos2}. At high oscillation frequencies, the
oscillating rectangular barrier can be replaced at leading order  by
its cycle-averaged (static) barrier, whose profile shows a
characteristic double-hump shape that sustains metastable states.
Thus resonant transmission of the incident particles across the
ac-driven barrier observed at certain below-barrier energies can be
simply explained as a resonant tunneling process, similar to e.g.
the Ramsauer-Townsend effect, the resonant electronic tunneling
across a double-barrier structure, or the resonant light
transmission of light in a Fabry-Perot optical cavity. In this work
we consider the coherent motion of a Bloch particle in a
tight-binding lattice with a strong impenetrable rectangular
potential barrier, and demonstrate that the application of an ac
field with appropriate amplitude and frequency can make the
potential barrier fully transparent. This phenomenon, which can be
again referred to as field-induced barrier transparency for Bloch
particles in analogy to its counterpart for free particles, has
however a very different origin. In particular, barrier transparency
is observed for incident Bloch waves at {\it any} allowed energy in
the band, i.e. it is not related to resonant tunneling of the
cycled-averaged potential as in the free particle case. Here we
investigate the phenomenon of field-induced barrier transparency of
Bloch waves by considering photonic transport in periodically-curved
waveguide lattices \cite{tras4,Longhi06}, however the present
analysis is rather generic to coherent transport in driven
tight-binding lattices and could be therefore of interest to other
physical set ups, such as to cold atoms or Bose-Einstein condensates
in accelerated optical lattices \cite{Arimondo}.\\
The paper is organized as follows. In Sec.II, the basic model is
described with specific reference to light transport in photonic
lattices, and a theoretical analysis of field-induced barrier
transparency in ac-driven tight-binding lattices is presented. In
Sec.III the theoretical predictions obtained in the high-frequency
regime are confirmed by direct numerical simulations, based on both
the time-periodic coupled-mode equations of the driven tight-binding
lattice and the Schr\"{o}dinger wave equation with a one-dimensional
potential. Deviations from the predictions based on the averaged
lattice model in the high frequency modulation limit, such as a
shift of the resonant curves, are briefly discussed. Finally, in
Sec.IV the main conclusions are outlined.

\section{Field-induced barrier-transparency in ac-driven tight-binding lattices: theoretical analysis}
 The starting point
of our analysis is provided by a standard model of light transport
in a one-dimensional array of tunneling-coupled optical waveguides
with a periodically-curved optical axis \cite{tras4,Longhi06}. In
the waveguide reference frame, light transport is described by the
following Schr\"{o}dinger-type wave equation for the electric field
envelope $\phi(x,z)$ (see, for instance, \cite{tras4,Longhi06bis})
 \begin{equation}
i \lambdabar \frac{\partial \phi}{\partial z} =
-\frac{\lambdabar^2}{2n_s} \frac{\partial^2 \phi}{\partial x^2} +
V(x) \phi-F(z)x \phi \equiv  \mathcal{H}_0 \phi-F(z)x \phi,
 \end{equation}
\begin{figure}
\includegraphics[scale=0.5]{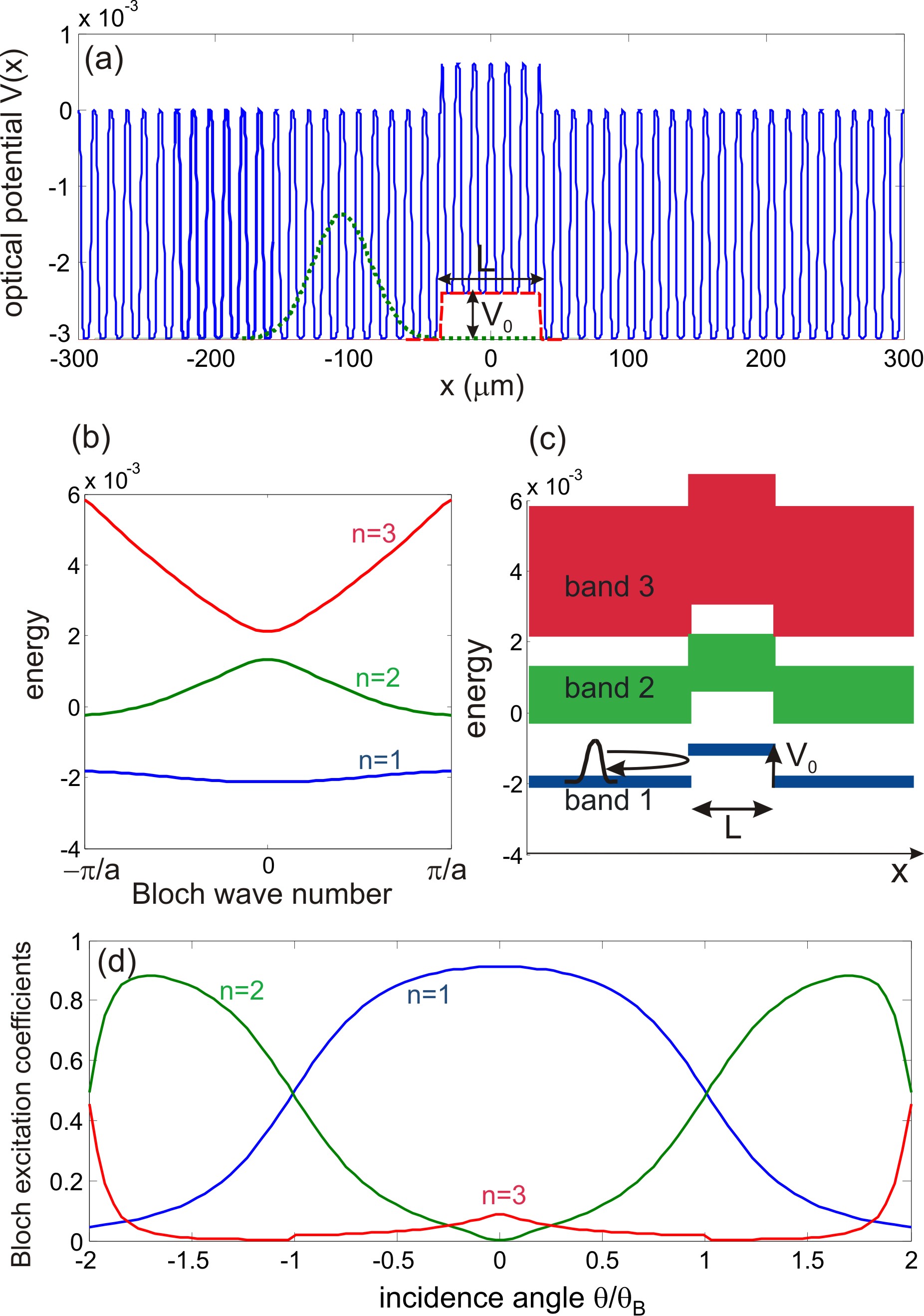}
\caption{ (a) Behavior of the optical potential $V(x)=n_s-n(x)$
(solid curve) of a one-dimensional lattice with a barrier, composed
by the superposition of the periodic optical potential (lattice
period $a$) and of the rectangular barrier potential (height $V_0$,
width $L$, represented by the dashed curve in the figure). The
dotted curve in the figure is the intensity profile of the incident
Gaussian wave packet in the numerical simulations of Sec.III.B. (b)
Band diagram (three lower-order bands) of the periodic optical
potential. The energy in the vertical axis is defined as the
eigenvalue of the Hamiltonian $\mathcal{H}_0$ of the periodic
potential entering in Eq.(1). (c) Space-energy band diagram of the
lattice with the rectangular barrier. The height $V_0$ of the
barrier is larger than the bandwidth $4 \Delta$ of the lowest
lattice band, but smaller than the gap separating the first two
bands. (d) Behavior of the Bloch excitation coefficients versus
incidence angle $\theta$ of a plane wave, normalized to the Bragg
angle $\theta_B= \lambda/(2a)$.}
\end{figure}
 where $\lambdabar=\lambda/(2 \pi)$ is the reduced wavelength of
 light waves, $n_s$ the substrate refractive index, $x$ and $z$ are the transverse and the
 longitudinal spatial coordinates, respectively, and $V(x)$ is the
 optical potential, which is related to the refractive index profile of the
straight array by the simple relation $V(x) \simeq n_s-n(x)$. The
last term on the right hand side of Eq.(1) is a fictitious
refractive index gradient arising from axis bending and  with a
$z$-varying slope given by $F(z)= -n_s \ddot{x}_0(z)$, where
$x_0(z)$ is the axis bending profile and the dot indicates the
derivatives with respect to $z$ \cite{tras4}. In its present form,
the paraxial wave equation (1) is formally equivalent to the
Schr\"{o}dinger equation describing the dynamics of a quantum
particle of mass $n_s$ in the potential $V(x)$ driven by a
time-dependent external force $F(z)$, provided that the spatial $z$
variable in the optical system is replaced by the temporal variable
in the quantum problem, and the photon wavelength $\lambda$ is
replaced by the Panck's constant. The potential $V$ is given by the
superposition of the periodic and barrier potentials, i.e.
$V(x)=V_p(x)+V_b(x)$, where $V_p(x+a)=V_p(x)$ is the periodic
potential describing a homogeneous array of equally-spaced wells,
and $V_b(x)$ is the barrier potential, which is assumed to describe
a rectangular barrier of height $V_0$ and width $L \gg a$ [see
Fig.1(a)]. The role of the barrier potential is to introduce a step,
of height $V_0$, into a finite number of wells in the lattice, say
from the waveguide $n=0$ to the waveguide $n=N$, as shown in
Fig.1(a). The potential height $V_0$ is typically assumed to be much
larger than the width of the lowest energy band of the array, yet
smaller than the gap between the first and second lattice bands, as
shown as an example in Fig.1(c). In this way, in the absence of the
ac driving force any Bloch wave packet, belonging to the lowest band
of the lattice and traveling along the lattice, is reflected from
the potential barrier [see Fig.1(c)]. To study the role of the ac
driving force and the possibility to make the barrier transparent,
let us introduce the nearest-neighboring tight-binding approximation
and let us assume that the lattice is mostly excited in its
lowest-order Bloch band. Such conditions are satisfied, for example,
by considering an array of weakly-coupled waveguides which is
initially excited by a broad beam tilted at an angle smaller than
the Bragg angle $\theta_B= \lambda/(2a)$
\cite{Longhi06,Longhi06bis}. Under such assumptions, from Eq. (1)
the following coupled-mode equations can be derived
\cite{tras4,Longhi06}
\begin{equation}
i\dot{c}_n=-\Delta(c_{n+1}+c_{n-1}) -f(z) n c_n +\sigma \rho_n c_n
\end{equation}
for the amplitudes $c_n$ of the field trapped in the individual
waveguides, where $\Delta > 0$ is the coupling constant between
adjacent waveguides,
\begin{equation}
f(z)\equiv
\frac{a}{\lambdabar}F(z)=-\frac{an_s}{\lambdabar}\ddot{x}_0(z)
\end{equation}
is the normalized forcing,
\begin{equation}
\sigma \equiv \frac{V_0}{\lambdabar},
\end{equation}
and
\begin{equation}
\rho_n \left\{
\begin{array}{cc}
1 & {\rm for} \; \; 0 \leq n \leq N \\
0 & {\rm otherwise} \; .
\end{array}
\right.
\end{equation}
 To study field-induced barrier transparency, it is worth
introducing, in place of $c_n$, the amplitudes $a_n$ defined by the
relations
\begin{equation}
a_n=c_n \exp \left[ -in \int_0^z d \xi f( \xi)+i \sigma \rho_n z
\right],
\end{equation}
 so that Eqs.(2) take the form
\begin{equation}
i\dot{a}_n=-\Delta_{n}(z) a_{n+1}-\Delta_{n-1}^*(z) a_{n-1}
\end{equation}
where we have set
\begin{equation}
\Delta_{n}(z)= \Delta \exp \left[ i \int_0^z d \xi f(\xi) +i\sigma
(\rho_{n}-\rho_{n+1}) \right].
\end{equation}
Let us now assume  a sinusoidal bending of waveguide axis with
spatial frequency $\omega= 2 \pi /\Lambda$ and amplitude $A$, i.e.
\begin{equation}
x_0(z)=A \cos(2 \pi z / \Lambda),
\end{equation}
 so that
\begin{equation}
 f(z)=f_0 \cos( 2 \pi z / \Lambda),
\end{equation}
 where
\begin{equation}
f_0=\frac{4 \pi^2 a n_s A}{ \Lambda^2 \lambdabar}.
\end{equation}
 Let us also assume that:\\
 (i) the barrier height $V_0$ and the quanta of
modulation $\lambdabar \omega$  are much larger than the width $4
\lambdabar \Delta$ of the tight-binding energy band;\\
(ii) the
modulation frequency is chosen such that the resonant condition
\begin{equation}
l \omega=\sigma=\frac{V_0}{ \lambdabar}
\end{equation}
is satisfied for some (small) integer $l$ (typically $l=1$ or
$l=2$). In the high-modulation frequency limit [assumption (i)], at
first order in a multiple-scale asymptotic analysis of Eqs.(7) the
spatial evolution of the amplitudes $a_n$  is dominated by the
cycle-average coupling rates
$\bar{\Delta}_n=(1/\Lambda)\int_0^{\Lambda}dz \Delta_n(z)$ (see, for
instance, \cite{LonghiPRB08}), i.e. one has
\begin{equation}
i\dot{a}_n \simeq -\bar{\Delta}_n a_{n+1}-\bar{\Delta}_{n-1}^*
a_{n-1}.
\end{equation}
Taking into account that $\exp[i \Gamma \sin (\omega z)]=\sum_n
J_n(\Gamma) \exp(i n \omega z)$ and using Eqs.(5) and (10), from
Eq.(8) it readily follows that
\begin{equation}
\bar{\Delta}_n= \left\{
\begin{array}{cc}
\Delta J_0(\Gamma) & {\rm for} \; \; n \neq -1,N \\
\Delta J_l(\Gamma) & {\rm for} \; \;  n=-1 \\
(-1)^l \Delta J_l(\Gamma) & {\rm for} \; \;  n=N
\end{array}
\right.
\end{equation}
 where we have set
\begin{equation}
\Gamma \equiv \frac{f_0}{\omega}=\frac{2 \pi a n_s A} {\lambdabar
\Lambda}.
\end{equation}
 Equations (13) thus describe
light transport in an effective lattice with two defects in the
coupling rates at lattice sites $n=-1$ and $n=N$. However, if the
amplitude $A$ of modulation is tuned such that
\begin{equation}
J_0(\Gamma)= \pm J_l(\Gamma),
\end{equation}
the effective lattice is homogeneous (i.e. defect-free), and thus
any Bloch wave packet propagates in the lattice without being
reflected \cite{note0}. That is, if the modulation frequency and
amplitude are tuned to satisfy the conditions (12) and (16), the
effect of the ac driving is to make the barrier $V_b$ fully
transparent for Bloch wave packets belonging to the lowest energy
band.
\begin{figure*}
\includegraphics[scale=1.1]{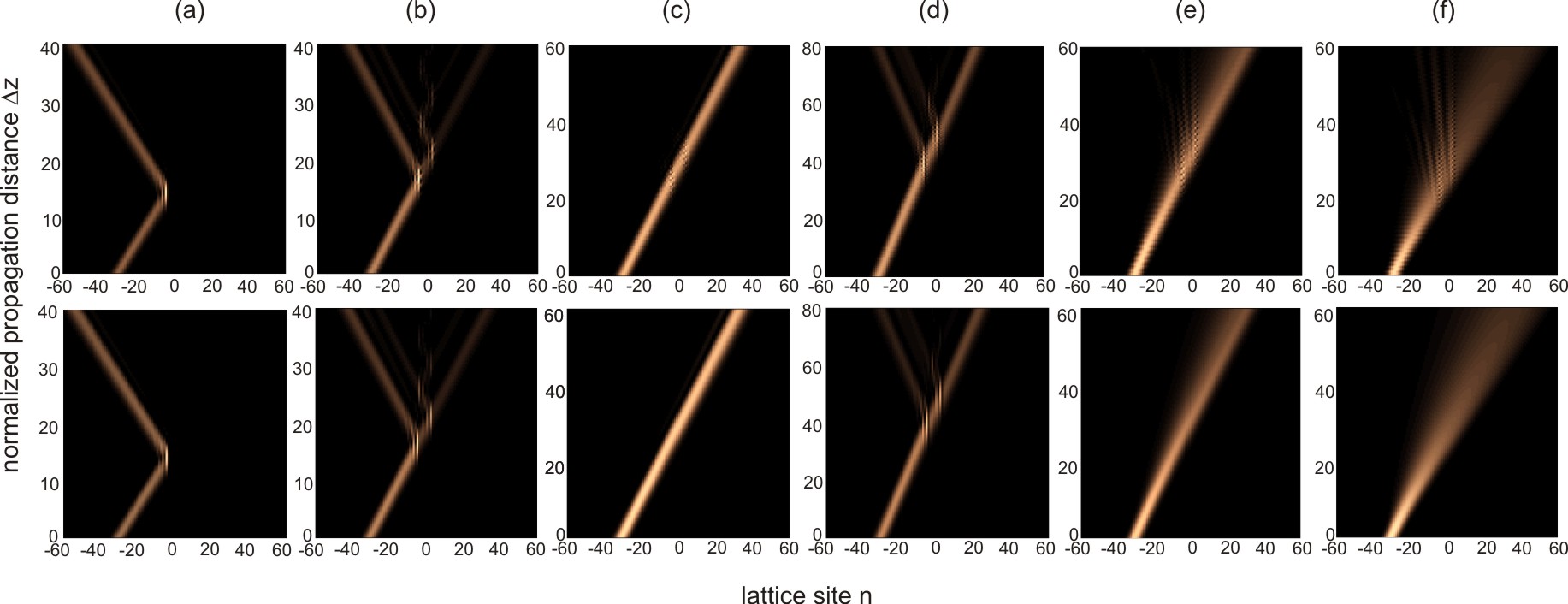}
\caption{Beam evolution (snapshots of lattice site intensities
$|c_n(z)|^2$)
 for an initial Gaussian wave packet distribution with transverse momentum $q$ as obtained by numerical simulations of the
 coupled-mode equations (2) (upper panels) and by the averaged equations (17) (lower
 panels) for $\sigma/ \Delta=4$ and $\omega=\sigma$. In (a-d), the initial transverse beam momentum is $q=\pi/2$, in (e) $q=\pi/3$, whereas in (f) $q=\pi/4$.
 In (a) $\Gamma=0$ (non-modulated lattice), in (b) $\Gamma=0.8$, in (c),(e),(f) $\Gamma=1.435$ (transparency condition), in (d) $\Gamma=1.8$.}
\end{figure*}

\begin{figure}
\includegraphics[scale=0.46]{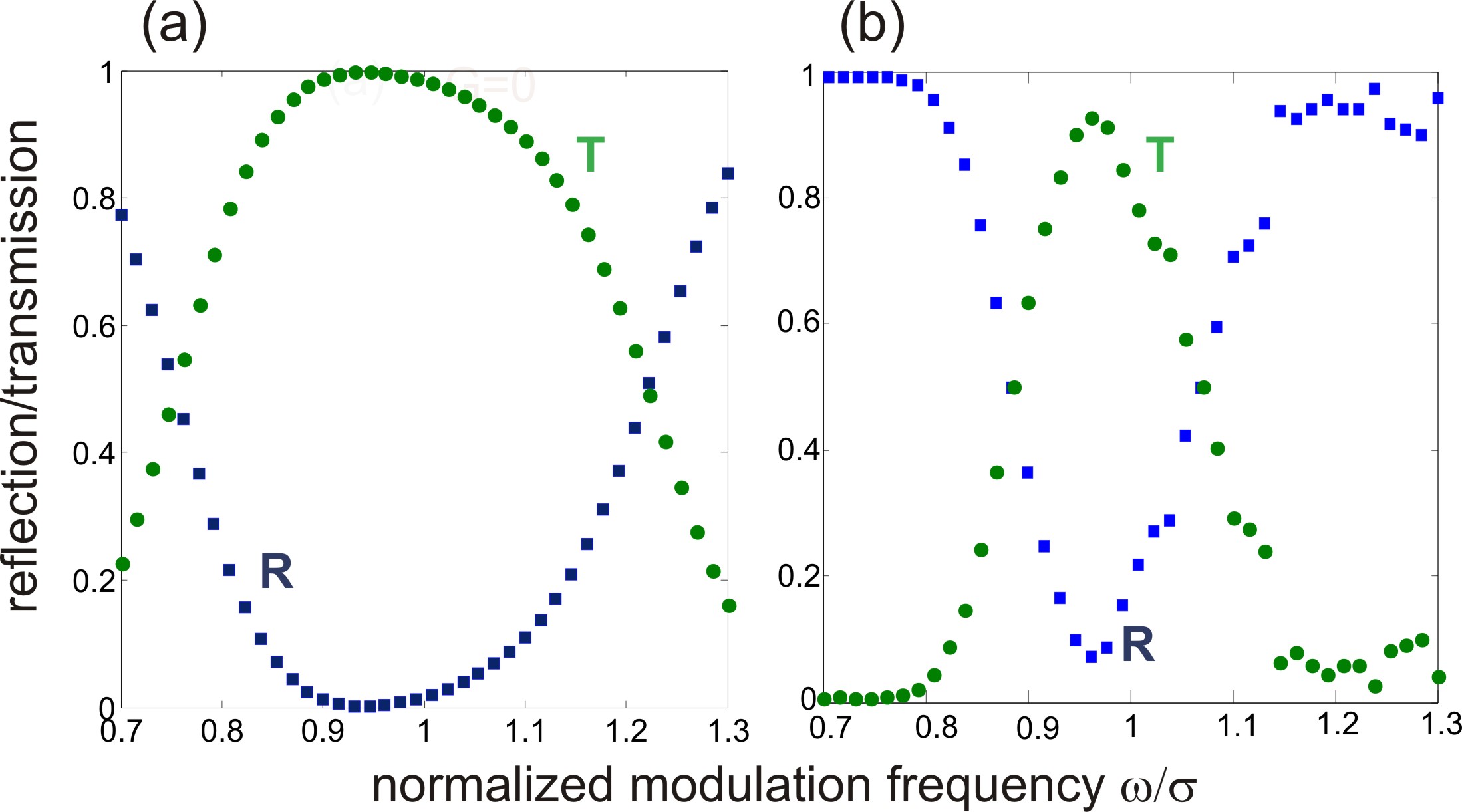}
\caption{Behavior of the power reflection ($R$) and transmission
($T$) coefficients of a Gaussian wave packet at the first barrier
discontinuity near the first transparency resonance $\omega=\sigma$
numerically computed (a) using the tight-binding model (2), and (b)
the full wave equation (1). Parameter values in the two cases are
given in the text.}
\end{figure}

\section{Numerical simulations}
In this section we confirm, by direct numerical simulations of both
the tight-binding equations (2) and the full wave equation (1), the
field-induced barrier transparency phenomenon predicted in the
previous section in the high-frequency modulation regime.

\subsection{Tight-binding lattice model}
Let us first consider beam reflection and transmission in the
tight-binding lattice model described by the coupled-mode equations
(2) with periodic coefficients. Equations (2) have been numerically
integrated using an accurate fourth-order variable-step Runge-Kutta
methods with absorbing boundary conditions. As an example,
Figs.2(a-d) (upper panels) show the numerically-computed evolution
of lattice site occupation probabilities $|c_n(z)|^2$ for
$\sigma/\Delta=4$, $N=6$, $\omega=\sigma$ and for a few increasing
values of the modulation amplitude, measured by the dimensionless
parameter $\Gamma=f_0/\omega$. For comparison, the lower panels in
Figs.2(a-d) show the corresponding evolution of occupation
probabilities as obtained by integration of the averaged effective
lattice equations [Eqs.(13) and (14)], valid in the high modulation
frequency limit. The array is initially excited by a Gaussian
distribution $c_n(0)=\exp[-(n-n_0)^2/w^2] \exp(iqn)$ with mean
position $n_0=-26$, width $w=4$ and momentum $q=\pi/2$
(corresponding to the largest cycle-averaged group velocity far from
the barrier). Note that at $\Gamma =1.435$, at which the condition
(16) is satisfied for $l=1$, the wave packet is not reflected, and
the barrier appears to be fully transparent [see Fig.2(c)]. In the
simulations shown in Fig.2, the ratio $\omega/\Delta=4$ is large
enough to ensure the validity of the averaged equations (13), at
least at first-order approximation. In fact, a more careful
comparison of the results obtained from the original coupled-mode
equations with periodic coefficients [Eqs.(2) or (7)] and the
averaged equations (13) shows some slight discrepancies, which
basically arise from neglecting higher-order terms in the asymptotic
analysis of Eqs.(7) (for more details see, for instance,
\cite{LonghiPRB08}). In particular, according to the average model
(13), transparency should be observed regardless of the initial
value of wave packet momentum $q$, i.e. of its mean energy; however,
numerical simulations of the original (periodic) coupled-mode
equations (2) show that some reflected light is observed at the
barrier when the initial wave packet momentum $q$ (i.e., its mean
group group velocity) is reduced. This is shown, as an example, in
Figs.2(e) and (f), where the evolution of lattice site occupation
probabilities are depicted for the same initial Gaussian wave packet
of Fig.2(c) (i.e. at the transparency condition), but with an
initial momentum lowered to $q=\pi/3$ [Fig.2(e)] and $q=\pi/4$
[Fig.2(f)]. Another difference between the average and periodic
coupled-mode equations can be seen by computing the first resonance
curve of the transparency process, depicted in Fig.3(a). The figure
shows the behavior of the power reflection $R$ and transmission $T$
coefficients versus $\omega$ (near the first resonance $\omega \sim
\sigma=V_0/ \lambdabar$) for the same initial Gaussian wave packet
with momentum $q=\pi/2$, as obtained by numerical integration of
Eqs.(2). For the sake of simplicity, the coefficients $R$ and $T$
have been computed by considering the first potential step solely of
the barrier, so that multiple reflections that would arise in the
presence of the two potential discontinuities of the rectangular
barrier [see, for example, Fig.2(b)] are avoided . For each
modulation frequency $\omega$, the modulation amplitude $A$ was
correspondingly varied such that the ratio $\Gamma=f_0/\omega$ [see
Eq.(15)] remains constant and equal to 1.435, at which the condition
(16) is satisfied. Note that, the condition of exact transparency
($R=0$) is attained at the ratio $\omega/ \sigma \simeq 0.95$, which
is slightly smaller than $1$, as expected from the averaged model
[Eq.(12) with $l=1$]. As discussed above, the validity of the
averaged model becomes more accurate, and the agreement between the
averaged and original (periodic) coupled mode equations closer, as
the ratio $\sim \sigma / \Delta$ of barrier height and width of
tight-binding energy band is increased.

\begin{figure}
\includegraphics[scale=0.6]{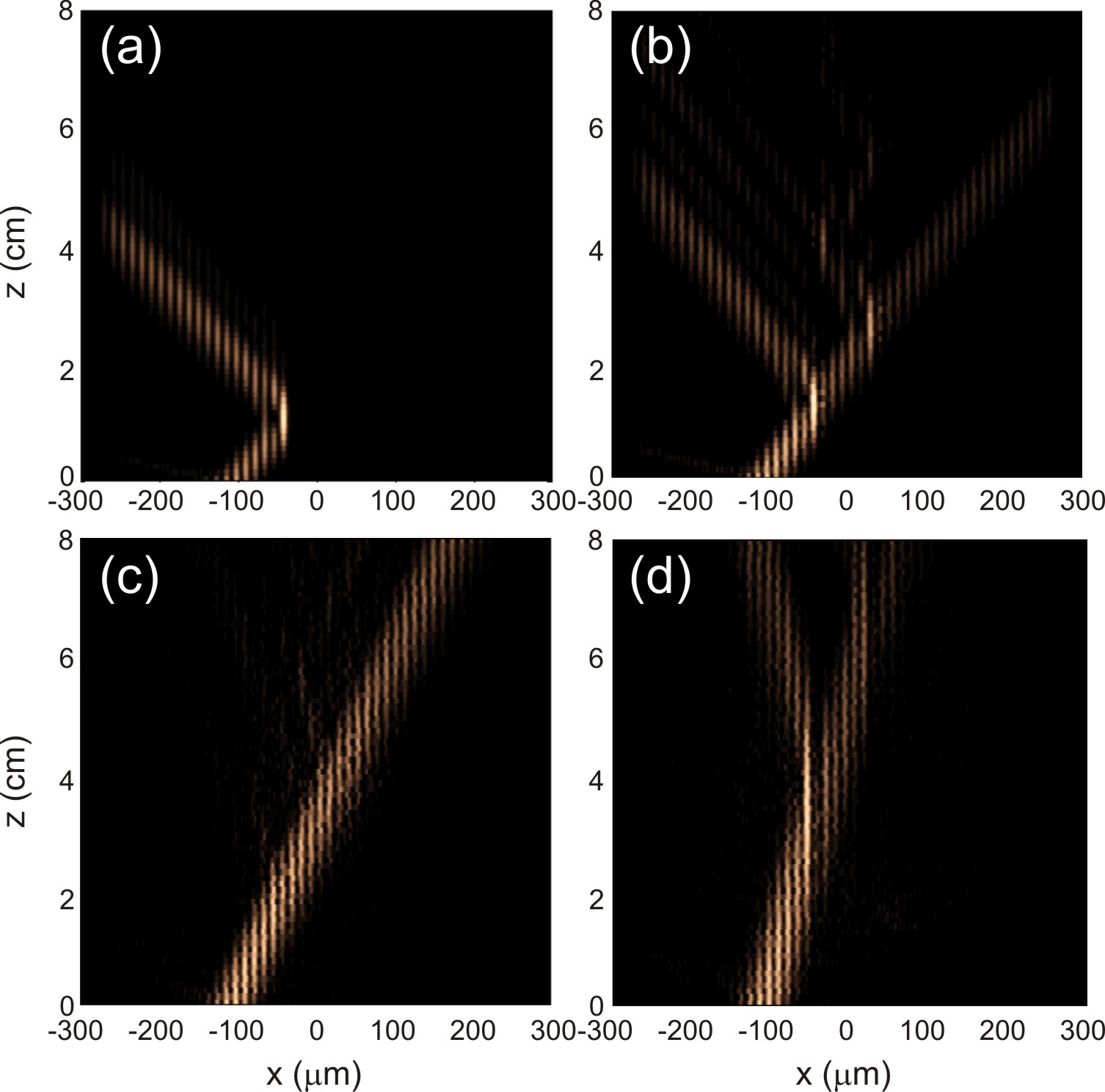}
\caption{ Beam intensity evolution (snapshot of $|\phi(x,z)|^2$) of
an initial Gaussian beam along the periodically-curved waveguide
array of Fig.1(a) for increasing values of
 the modulation amplitude [measured by the parameter $\Gamma$, given by Eq.(15)]: (a) $\Gamma=0$ (straight array); (b) $\Gamma=0.8$,
 (c) $\Gamma=1.435$ (transparency condition), and
 (d) $\Gamma=1.8$. In all the simulations, the input Gaussian beam is tilted at the angle $\theta=\theta_B/2$, and the modulation frequency is $\omega=0.98 \sigma$.}
\end{figure}

\begin{figure}
\includegraphics[scale=0.43]{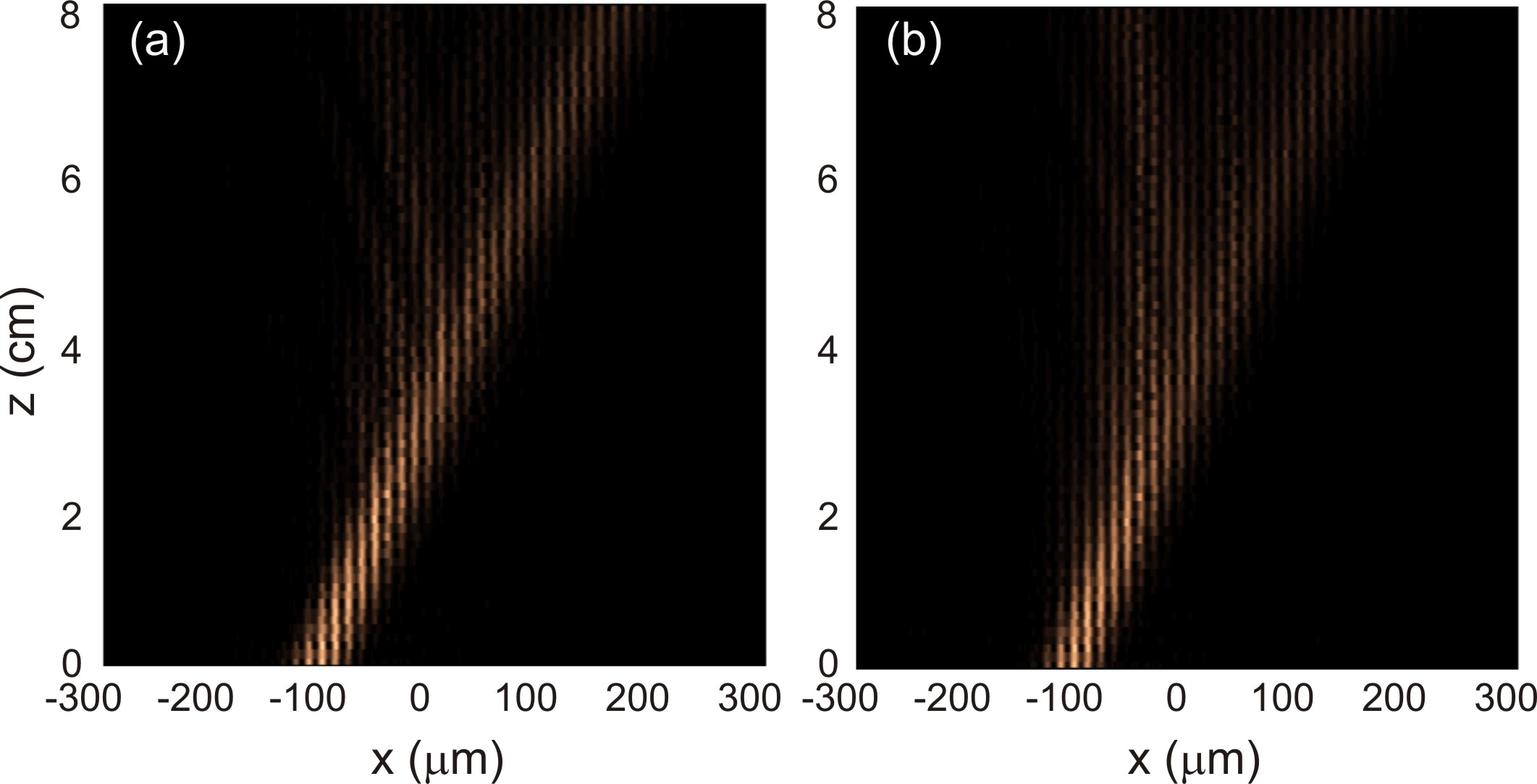}
\caption{ Same as Fig.4, but for an initial Gaussian beam with
$\Gamma=1.435$, $\omega=0.98 \sigma$ and tilt angle (a)
$\theta=\theta_B/3$, and (b) $\theta=\theta_B/4$.}
\end{figure}

\subsection{Full-wave equation} We checked the validity of the
tight-binding lattice analysis and the onset of field-induced
barrier transparency in photonic waveguide arrays by direct
numerical simulations of the paraxial wave equation (1) using
standard pseudospectral methods. The optical potential
$V(x)=n_s-n(x)$ of the lattice used in numerical simulations is
shown in Fig. 1(a) and corresponds to a typical effective index
profile of Lithium-Niobate waveguide arrays, fabricated by the
proton-exchange technique and probed at $\lambda=1.44 \;\mu$m
wavelength (see, for instance, the experiment reported in
Ref.\cite{Longhi06bis}). The refractive index profile $n_w(x)$ of
each waveguide in the array is taken to be given by
\cite{Longhi06bis}
\begin{equation}
n_w(x)=\Delta n \frac{{\rm erf}[(x+w_g)/D_x]-{\rm
erf}[(x-w_g)/D_x]}{2{\rm erf}(w_g/D_x)}
\end{equation}
where $w_g=3.5 \; \mu$m is the channel width, $D_x=1 \; \mu$m the
diffusion length, and $\Delta n=0.003$ the peak index change. The
waveguide spacing (lattice period) is $a=12 \; \mu$m, and the
substrate refractive index at the probing wavelength is
$n_s=2.1381$. The band structure of the lattice, numerically
computed by a standard plane-wave expansion method, is shown in
Fig.1(b). From the width of the lowest order (tight-binding) lattice
band, a coupling rate $\Delta \simeq 3.25 \; {\rm cm}^{-1}$ can be
estimated between adjacent waveguides. When the lattice is excited
by a broad wave packet, tilted at an angle $\theta$, it generally
breaks up into the superposition of different wave packets,
belonging to the various lattice bands and refracting at different
angles (see, for instance, \cite{Longhi06bis,Blochexcitation}). The
fractional excitations of the different lattice bands are given by
the Bloch-wave excitation coefficients $C_n$, defined as in
Ref.\cite{Blochexcitation}. Figure 1(d) shows the
numerically-computed behavior of $C_n(\theta)$ for the various
lattice bands of the periodic part of the optical potential of
Fig.1(a) versus the tilt angle $\theta$ of the incident beam, in
units of the Bragg angle $\theta_B= \lambda/(2a)$. As one can see,
for broad input beams tilted at an angle $\theta$ smaller than half
of the Bragg angle, the lowest-order lattice band is mainly excited,
which indicates that in this case the tight-binding model of Sec.II
can be safely applied. The barrier height used in the numerical
simulations is $V_0=\Delta n/5=6 \times 10^{-4}$, whereas its width
is $L=72 \; \mu$m. A schematic of the space-dependent band diagram
of the lattice, shown in Fig.1(c), clearly indicates that, in the
absence of the external ac field, the barrier is impenetrable and
any Bloch wave packet, incident onto the barrier, is reflected. This
is shown, as an example, in Fig.4(a), where the numerically-computed
evolution of the field intensity $|\phi(x,z)|^2$ along the
non-modulated lattice is depicted by assuming, as an initial
condition, the tilted Gaussian beam $\phi(x,0)=\exp[-(x+x_0)^2/w^2]
\exp(2 \pi i  \theta x /\lambda)$ with spot size $w=30 \; \mu$m,
offset $x_0=108 \; \mu$m [see the dotted curve in Fig.1(a)] and tilt
angle $\theta=\theta_B/2 \simeq 1.719^{\rm o}$. Let us now introduce
a sinusoidal modulation of the axis bending $x_0(z)$ at a spatial
frequency $\omega=2 \pi /\Lambda$ close to the first resonance
[$l=1$ in Eq.(12)], which corresponds to a spatial modulation period
$\Lambda=2.45$ mm. Note that the ratio between modulation frequency
$\omega$ and coupling rate $\Delta$ of waveguides turns out to be
$\sim 7.9$, i.e. the high frequency modulation condition is well
satisfied. Figures 4(b-d) show the evolution of the same Gaussian
wave packet as in Fig.4(a) but in the periodically-curved waveguide
array for increasing values of the modulation amplitude $A$,
measured by the dimensionless parameter $\Gamma$ given by Eq.(15).
Note that the cycle-averaged transverse group velocity of the wave
packet, far from the barrier region and at the tilting angle
$\theta=\theta_B/2$, is equal to $v_g=2\Delta a |J_0(\Gamma)|$, and
thus decreases as $\Gamma$ is increased from zero [Fig.4(a)] to
$\Gamma=1.8$ [Fig.4(d)]. As the modulation is increased, the
Gaussian wave packet is less and less reflected from the barrier,
till a nearly reflectionless regime, corresponding to full barrier
transparency, is attained when the transparency condition (16) is
reached [see Fig.4(c)]. Barrier transparency is observed at
different values of the initial tilt angle $\theta$ (i.e. initial
wave packet momentum), expect for small tilt angles at which the
wave packet shows a more complex dynamics at the barrier crossing
(see Fig.5). This behavior, already noticed in Sec.II in the
framework of the tight-binding analysis, is mainly ascribable to a
discrepancy between the average model [Eq.(13)] and the original
coupled-mode equations with periodic coefficients [Eqs.(2) or (7)].
A slight discrepancy can be  also seen when computing the resonance
curve of the field-induced barrier transparency process using the
full wave equation (1), similarly to what already noticed within the
tight-binding model [see Fig.3(a)]. In Fig.3(b), the
numerically-computed power reflection ($R$) and transmission ($T$)
coefficients versus $\omega$ (near the first resonance $\omega \sim
\sigma=V_0/ \lambdabar$) are depicted for the Gaussian wave packet
of Fig.4. Like for the resonance curve computed using the
tight-binding model [Fig.3(a)], the coefficients $R$ and $T$ have
been calculated by considering the first potential step solely of
the barrier, and for each modulation frequency $\omega$ the
modulation amplitude $A$ was correspondingly varied such that
$\Gamma=f_0/\omega=1.435$. Note that, as in Fig.3(a), the resonance
frequency turns out to be slightly smaller than the theoretical
value $\sigma$ predicted by the averaged model.\\
As a concluding remark, it is important to stress that, as compared
to the phenomenon of field-induced barrier transparency for a
freely-moving quantum particle \cite{f1} in which particle
transmission occurs solely at special values of initial momentum
(energy) that match the metastable states of the cycle-average
potential barrier, the phenomenon of barrier transparency for Bloch
wave packets predicted in this work is relatively insensitive to the
initial wave packet momentum and can not thus be ascribed to a
resonant tunneling phenomenon as for the free particle. However, as
opposed to the case of Ref.\cite{f1}, where the frequency and
amplitude of the ac field may take relatively arbitrary values
\cite{note}, in our system the modulation frequency as well as the
modulation amplitude should satisfy certain resonance conditions,
namely Eqs.(12) and (16).

\section{Conclusions}
In this work it has been theoretically shown that a rectangular
potential barrier for a Bloch particle in a tight-binding lattice
can be made fully transparent by the application of a strong ac
field with appropriate amplitude and frequency. As this phenomenon
bears some connection with the field-induced barrier transparency
phenomenon previously predicted for freely-moving quantum particles
scattered by an ac-driven potential barrier \cite{f1,f0}, the
transparency effect for the Bloch particle has a rather different
physical origin, as discussed in this work. In particular, in the
high-frequency limit transparency is attained independently of the
energy of the Bloch wave packet, and therefore particle transmission
across the barrier can not be explained as a resonant tunneling
process across the cycle-averaged potential, as for a freely moving
particle \cite{f1}. This phenomenon could be experimentally observed
in periodically-curved waveguide arrays, where light transport along
the lattice mimics the coherent temporal evolution of a Bloch
particle with an external ac driving field \cite{Longhi06}.
Numerical simulations based on the paraxial (Schr\"{o}dinger) wave
equation have also shown that the phenomenon of field-induced
barrier transparency persists beyond the tight-binding
approximation, provided that the lattice is initially excited by a
broad beam tilted at an angle smaller than the Bragg angle.

 \acknowledgments
 This work was supported by the Italian MIUR (Grant
No. PRIN-20082YCAAK, "Analogie ottico-quantistiche in strutture
fotoniche a guida d'onda").


\begin{thebibliography}{}
\bibitem{note0}
It should be noted that, once Eq.(16) is satisfed, a change of the
sign of the coupling rate between adjacent sites at $n=-1$ or $n=N$
can occur. However, such a sign change  does not introduce
reflection of a wave packet propagating in the lattice. In fact, by
a suitable change of the phase of the amplitudes $a_n$, the lattice
(13) with some sign changes in the coupling rates become fully
equivalent to a homogeneous lattice with constant coupling rates
between adjacent sites.
\bibitem{note}
The only condition on the modulation frequency is that its value
should be large enough in order to ensure the validity of the
average potential model. The modulation amplitude should be also
large enough to ensure that the cycle-averaged potential barrier
sustains some metastable states (see \cite{f1} for more details).
\end{thebibliography}
\end{document}